\documentclass[fleqn,usenatbib]{mnras}

\usepackage[T1]{fontenc}
\usepackage{ae,aecompl}

\usepackage{eso-pic}

\AddToShipoutPictureBG*{%
  \AtPageUpperLeft{%
    \hspace{0.75\paperwidth}%
    \raisebox{-4.5\baselineskip}{%
}}}%

\usepackage{graphicx}	
\usepackage{amsmath}	
\usepackage{amssymb}	
\usepackage[shortlabels]{enumitem}

\usepackage{multicol}
\usepackage{caption}
\usepackage{subcaption}

\usepackage{newtxtext,newtxmath}

\newcommand{\mulim}{$\mu_{lim}$}

\defcitealias{Martin2022}{M22}

\title[Tidal feature identification with CNNs]{Identification of tidal features in deep optical galaxy images with Convolutional Neural Networks}

\author[ Dom\'inguez S\'anchez et al.]{
\parbox{\textwidth}{
\Large
H.~Dom\'{i}nguez S{\'a}nchez$^{1, 2}$\thanks{Corresponding author: \texttt{\rm \texttt{hdominguez@cefca.es}}}, 
G.~Martin$^{3, 4}$,  
I.~Damjanov$^{5}$,
F.~Buitrago$^{6, 7}$,
M.~Huertas-Company$^{8, 9, 10}$,
C. Bottrell$^{11}$,
M.~Bernardi$^{12}$, 
J. H.~Knapen$^{8, 9}$,
J.~Vega-Ferrero$^{6, 8, 9}$,
R.~Hausen$^{13}$,
E.~Kado-Fong$^{14}$,
D.~Población-Criado$^{15}$,
H.~Souchereau$^{5, 16}$,
O.~K. Leste$^{17}$,
B.~Robertson$^{18}$,
B.~Sahelices$^{15}$,
K.V.~Johnston$^{19}$
  }
  \vspace{0.4cm}\\~\\
$^{1}$ Centro de Estudios de Física del Cosmos de Aragón (CEFCA), Plaza San Juan, 1, 44001, Teruel, Spain\\
$^{2}$ Institute of Space Sciences (ICE, CSIC), Campus UAB, Carrer de Can Magrans, s/n, 08193 Barcelona, Spain\\
$^{3}$ Korea Astronomy and Space Science Institute, 776 Daedeokdae-ro, Yuseong-gu, Daejeon 34055, Korea\\
$^{4}$ Steward Observatory, University of Arizona, 933 N. Cherry Ave, Tucson, AZ 85719, USA\\
$^{5}$ Department of Astronomy and Physics, Saint Mary's University, 923 Robie Street, Halifax, NS B3H 3C3, Canada \\
$^{6}$ Departamento de F\'{i}sica Te\'{o}rica, At\'{o}mica y \'{O}ptica, Universidad de Valladolid, 47011 Valladolid, Spain\\
$^{7}$ Instituto de Astrof\'{\i}sica e Ci\^{e}ncias do Espa\c{c}o, Universidade de Lisboa, OAL, Tapada da Ajuda, PT1349-018 Lisbon, Portugal\\
$^{8}$ Instituto de Astrofísica de Canarias (IAC), La Laguna, 38205, Spain\\
$^{9}$ Departamento de F\'isica - Universidad de La Laguna, La Laguna, 38205, Spain\\
$^{10}$ LERMA - Observatoire de Paris, PSL, Universit\'e Paris-Cit\'e, Paris, France\\
$^{11}$ Kavli Institute for the Physics and Mathematics of the Universe (WPI), UTIAS, University of Tokyo, Kashiwa, Chiba 277-8583, Japan \\
$^{12}$ Department of Physics and Astronomy, University of Pennsylvania, Philadelphia, PA 19104, USA \\
$^{13}$Department of Physics and Astronomy, The Johns Hopkins University, 3400 N. Charles St., Baltimore, MD 21218 USA\\
$^{14}$ Physics Department, Yale Center for Astronomy \& Astrophysics, PO Box 208120, New Haven, CT 06520, USA\\
$^{15}$ GCME Research Group, Departamento de Informática, Universidad de Valladolid, 47011 Valladolid, Spain\\
$^{16}$ Department of Astronomy, Yale University, New Haven, CT 06511, USA\\
$^{17}$ Department of Physics and Astronomy, University of Victoria, 3800 Finnerty Rd, Victoria, BC V8P 5C2, Canada\\
$^{18}$ Department of Astronomy and Astrophysics, University of California, Santa Cruz, 1156 High Street, Santa Cruz, CA 95064 USA\\
$^{19}$ Department of Astronomy, Columbia University, New York, NY, USA\\
\vspace{-1cm}
}

\pubyear{2023}

\begin{document}
\label{firstpage}
\pagerange{\pageref{firstpage}--\pageref{lastpage}}
\maketitle

\begin{abstract}

Interactions between galaxies leave distinguishable imprints in the form of tidal features which hold important clues about their mass assembly. Unfortunately, these structures are difficult to detect because they are low surface brightness features so deep observations are needed. Upcoming surveys promise several orders of magnitude increase in depth and sky coverage, for which automated methods for tidal feature detection will become mandatory. We test the ability of a convolutional neural network to reproduce human visual classifications for tidal detections. We use as training $\sim$6000 simulated images classified by professional astronomers. The mock Hyper Suprime Cam Subaru (HSC) images  include variations with redshift, projection angle and  surface brightness (\mulim =26~-~35 mag arcsec$^{-2}$). We obtain satisfactory results with accuracy, precision and recall values of Acc=0.84,  P=0.72 and R=0.85 for the test sample. While the accuracy and precision values are roughly constant for all surface brightness,  the recall (completeness) is significantly affected by image depth. The recovery rate shows strong dependence on the type of tidal features: we recover all the images showing \textit{shell} features and 87\% of the tidal \textit{streams}; these fractions are below 75\% for \textit{mergers, tidal tails} and \textit{bridges}. When applied to real HSC images, the performance of the model worsens significantly. We speculate that this is due to the lack of realism of the simulations and take it as a warning on applying deep learning models to different data domains without prior testing on the actual data.

\end{abstract}




\section{Introduction}

In the standard $\Lambda$-Cold Dark Matter ($\Lambda$CDM) cosmology scenario, small scale overdense perturbations  in the early Universe collapse first and produce dark matter haloes that accumulate baryons at the centre. The small structures aggregate successively into larger structures via mergers in a process known as hierarchical growth \citep{White1978, Fall1980, White1991, Lacey1993}. In addition, accretion processes of small satellite galaxies or gas in filaments produce a vast and complex network of ultra-low surface brightness streams which should be present around all galaxies (e.g., \citealt{Pillepich2014} and references therein).

Therefore, galaxy mergers have a fundamental and critical role within the $\Lambda$CDM cosmogony. While there is a general consensus that the merger fraction increases with galaxy stellar mass, both from simulations (e.g., \citealt{RodriguezGomez2016, Husko2022}) and observations \citep{vanDokkum2010, CLS2012, RodriguezPuebla2017}, the relative contribution of in-situ star formation and  accreted stellar mass remains an open question across much of the galaxy mass spectrum (e.g., \citealt{Qu2017, Fitts2018, Conselice2022}). 
The rate of major and minor merger events, and their impact on galaxy mass assembly and morphological transformations, are also under debate (e.g. \citealt{Lotz2011, Lofthouse2017, Martin2017, Martin2018, Martin2021}).

Minor mergers (with baryonic mass ratios below 1:4) are expected to be significantly more common than major ones (e.g., \citealt{Cole2000, Lotz2011}), and to remain frequent even at the present epoch (although this is still under debate, see for example \citealt{OLeary2021}). As minor mergers do not necessarily destroy pre-existing stellar disks (e.g., \citealt{Robertson2006}), signs of recent or ongoing minor mergers should be apparent around galaxies in the form of stellar tidal features, which extend into the halo of the central galaxy. Merger remnants which are only a few dynamical periods old should leave distinguishable imprints in the outskirts of galaxies.  The frequency and characteristics of these  features hold vital clues to the nature of the events which have created them \citep{Hernquist1989, Mihos1998, Helmi1999, MartinezDelgado2009, Hendel2015, Spavone2020, Montes2020, VeraCasanova2022} and can thus be used to disentangle the different formation channels. Following \cite{Duc2015}, \textit{tails} are expected to be pulled out material from a gas-rich disc galaxy, while \textit{streams} would be stripped material from a low-mass companion being consumed by the primary galaxy. Other features such as \textit{fans} and \textit{plumes} are expected to come from dry, major mergers.
In addition, \textit{clouds} and \textit{shells} are expected to be the result of interactions with radial orbits, while \textit{great circles} are more predominant for circular orbits events \citep{Johnston2008}. 

Unfortunately, the majority of tidal features have very low surface brightness, expected to be fainter than 29 mag arcsec$^{-2}$ in the $r$-band \citep{Bullock2005, Cooper2013},  and extremely deep observations are necessary to detect them, as shown explicitly in \citet{Conselice2000, Ji2014, Bottrell2019, Thorp2021} and  \cite{Mancillas2019}, where the authors find two and three times more streams based on a surface brightness cut of 33 mag arcsec$^{-2}$ than with 29 mag arcsec$^{-2}$.  Although there is an increasing interest in the literature on the identification and characterization of tidal features, most works focus on the detailed analysis of a small number of objects via visual inspection  (e.g., \citealt{MartinezDelgado2010, Javanmardi2016, Morales2018, MartinDelgado2021, Valenzuela2022, Huang2022, Sola2022}), some of them belonging to local groups or clusters of galaxies (e.g., \citealt{Iodice2017, Mihos2017, Spavone2018}). A sample of 92 ETGs galaxies from ATLAS$^{\rm 3D}$ was presented in \cite{Duc2015}, reporting signs of interactions or perturbed morphologies in more than half of them, thanks to an observing strategy and data reduction pipeline  optimized for low surface brightness features. \cite{Hood2018} present a visual identification of galaxies with tidal features based on DECam Legacy Survey images ($r$-band 3$\sigma$ depth of $\sim$27.9 mag arcsec$^{-2}$) but due to the small area inspected (100 arcsec$^{2}$), less than 200  of them have tidal features detected with high confidence.
One of the largest catalogues of tidal detections up to date  was  presented in \cite{KadoFong2018} using The Hyper Suprime-Cam Subaru Strategic Program (HSC-SSP,  \citealt{Miyazaki2012}) data. The authors applied applied a filtering algorithm  that iteratively separates low and high spatial frequency features of   images, resulting in a  sample of $\sim$1200 galaxies with tidal detections from a sample of $\sim$20,000.

With the arrival of large imaging surveys such as Euclid \citep{Laureijs2011} and the Vera Rubin Observatory's Legacy Survey of Space and Time (LSST, \citealt{Ivezic2019A}), the detection of these features via visual inspection is unfeasible and automated methods become imperative. The use of supervised deep learning for the analysis of galaxy images, such as convolutional neural networks (CNN), has proven to be extremely successful for classifying galaxy images (e.g, \citealt{Dieleman2015, Huertas2015, DS2022, Cheng2020, VegaFerrero2021, Walmsley2022, Hausen2020, Ghosh2020}), including classifications of relatively rare objects such as strong lensed galaxies \citep{Lanusse2018, Cheng2020} or post-mergers (e.g., \citealt{Bickley2021}). However, one of the main drawbacks of supervised learning approaches is the need for a large sample of labelled galaxies (of the order of thousands) to train and test the algorithm and its performance in different regimes (see \citealt{HuertasCompany2022} for a review on the topic). An alternative is the use of simulations: the viability of using galaxies from hydrodynamical simulations to train deep learning models to classify real galaxies and mergers has indeed been shown in \cite{Bottrell2019b} and \cite{HuertasCompany2019}. The scarcity of a large number of galaxies showing tidal features to be used as training data has prevented to develop automated supervised detection algorithms and so far there have been almost no attempts in the literature to this respect. A pioneering effort to develop a CNN for tidal stream detection was presented in \citet{Walmsley2019}, where the authors used imaging for the Canada–France–Hawaii Telescope Legacy Survey-Wide Survey \citep{Gwyn2012}. However, the models only achieved a 76 percent accuracy, probably due to the small size of the training sample ($\sim$~1700 galaxies,  of which only 305 showed tidal stream detections).

In this work, we use synthetic HSC images of galaxies generated by the \textsc{Newhorizon} cosmological simulations \citep{Dubois2021} to examine the viability of using CNNs to identify galaxies with tidal features. The original sample, described in Section \ref{sect:data}, includes $\sim$~6000 images at different surface brightness limits classified by professional astronomers. This is the largest catalogue of tidal features based on visual classification up to date. We describe our CNN and training strategy in  Section \ref{sect:cnn} and test the ability of the CNNs to recover human-like classifications in Section \ref{sect:results}, where we present the performance of the model as a function of the feature class (Section \ref{sect:class}), redshift (Section \ref{sect:z}) and image depth (Section \ref{sect:SB}).  The outcome of applying the models to real data are discussed in Section \ref{sect:real-data}, including an attempt of using the \cite{KadoFong2018} classification as a training sample. We  summarise our results and discuss their implications  in Section  \ref{sect:summary}.

\begin{figure*}
  \centering
  \begin{tabular}{c c c c c}
  \text{\mulim = 28 mag arcsec$^{-2}$} & \text{\mulim = 29 mag arcsec$^{-2}$} &  \text{\mulim = 30 mag arcsec$^{-2}$} &  \text{\mulim = 31 mag arcsec$^{-2}$} &  \text{\mulim = 35 mag arcsec$^{-2}$} \\
  \includegraphics[width=0.176\linewidth]{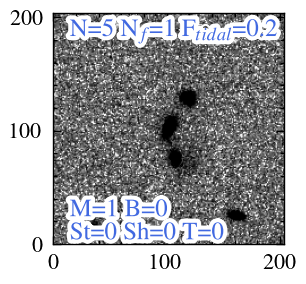}&
  \includegraphics[width=0.176\linewidth]{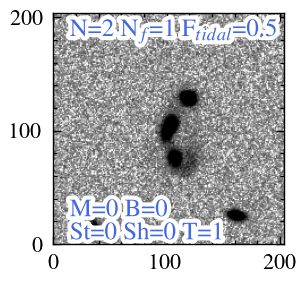} &
  \includegraphics[width=0.176\linewidth]{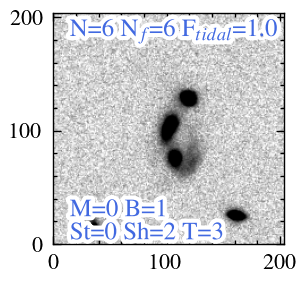} &
  \includegraphics[width=0.176\linewidth]{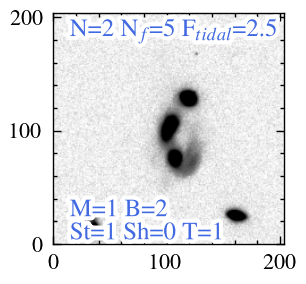} &
  \includegraphics[width=0.195\linewidth]{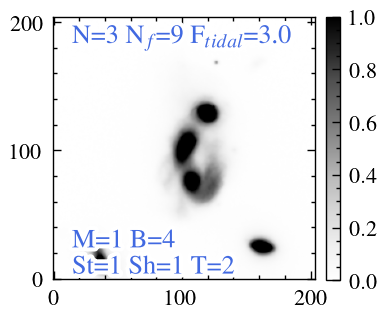} 
  \end{tabular}
  \caption{Example of the classification performed in \citetalias{Martin2022}. Images of the same galaxy observed at z=0.05 with different \mulim (28, 29, 30, 31, 35 mag arcsec$^{-2}$, form left to right) were classified by a varying number of astronomers (N) into different categories. The number of observed features of each class is reported in the cutouts (St=streams, Sh=shells, T=tails, M=mergers, B=bridges) along with the total number of features (N$_f$) and F$_{\rm Tidal}$=N$_f$/N. In this work we consider as positive examples those images with F$_{\rm Tidal}$> 1, negative those with F$_{\rm Tidal}$=0 and uncertain otherwise. Following this criteria, the images with \mulim=28 and 29 have an uncertain classification, while those with \mulim > 29 are classified as showing tidal features. The cutouts are normalized in the (0,1) range using \textit{arcsinh stretch}, as described in Section \ref{sect:img}.}
  \label{fig:labels}
\end{figure*}


\section{Data}
\label{sect:data}

We take advantage of the galaxy images and labelling presented in \citet[herafter \citetalias{Martin2022}]{Martin2022}. The galaxies are generated with \textsc{Newhorizon}, state-of-the-art cosmological hydrodynamical simulations \citep{Dubois2021}, a zoom-in  of the parent \textsc{Horizon-AGN} simulation \citep{Dubois2014}. \textsc{Newhorizon} combines high stellar mass (1.3$\times$10$^4$ M$_{\odot}$) and spatial resolution ($\sim$ 34 pc) with a contiguous volume of (16 Mpc)$^3$. Given the diffuse nature of galaxy stellar haloes, the trade off between resolution and volume is an important consideration. The simulations adopt the cosmological parameters from \citet[$\Omega_{\rm m}=0.272$, $\Omega_\Lambda=0.728$, $\Omega_{\rm b}=0.045$, $H_0=70.4 \ \rm km\,s^{-1}\, Mpc^{-1}$]{Komatsu2011}.  We refer the reader to \citetalias{Martin2022} for more technical details on the simulations.

\subsection{Mock galaxy images}
\label{sect:mock-images}

The parent sample consists of 36 unique galaxies,  with masses  above 10$^{9.5}$M$_{\odot}$ at z=0.2, and their progenitors at z=0.4, 0.6 and 0.8. Realistic HSC-like  mock images are generated by convolving the smoothed star-particle fluxes with the $g$-band HSC 1D PSF \citep{Montes2021}. Three projections of each snapshot ($xy$, $xz$, $yz$) are created at 5 different distances (corresponding at z=0.05, 0.1, 0.2, 0.4 and 0.8). The physical field of view is 100 kpc (proper) cropped from the initial 1 Mpc cube. Mock images are produced for each galaxy by extracting star particles centred around each galaxy. The spectral energy distribution (SED) for each star particle is then calculated
from a grid of \cite{BC2003} simple stellar population models assuming a \cite{Salpeter1955} IMF. They account for dust attenuation of the SEDs using a screen model in front of each particle
for which  a gas-to-dust ratio of 0.4 \cite{Draine2007} and a \cite{Weingartner2001} R = 3.1 Milky Way dust grain model are assumed. After redshifting each particle SED, the flux of each
particle is calculated by convolving with the appropriate bandpass transmission function. Finally, random Gaussian noise is added to the simulated images to reach different limiting surface brightness $\mu_{r}^{lim}$  corresponding to 28, 29, 30, 31 and 35 mag arcsec$^{-2}$ (3$\sigma$, 10"$\times$10"). The combination of these parameters results in 10800 unique simulated images. Since the pixel angular size is fixed, the difference in distance of the galaxies directly translates into cutouts of different sizes (26$\times$26, 36$\times$36, 60$\times$60, 108$\times$108, 204$\times$204 pixels, from z=0.8 to z=0.05). In order to increase the sample size and to have mock images which resemble current observations better, we have generated 2$\times$1453 additional snapshots with \mulim = 26 and 27 mag arcsec$^{-2}$  by adding Gaussian noise  following Equation 3 from \citetalias{Martin2022} to the deepest available image of each particular counterpart (i.e., with the corresponding ID, snapshot, redshift and projection).

\subsection{Tidal feature classification}
\label{sect:cat-labels}

\begin{figure*}
  \centering
  \includegraphics[width=0.99\linewidth]{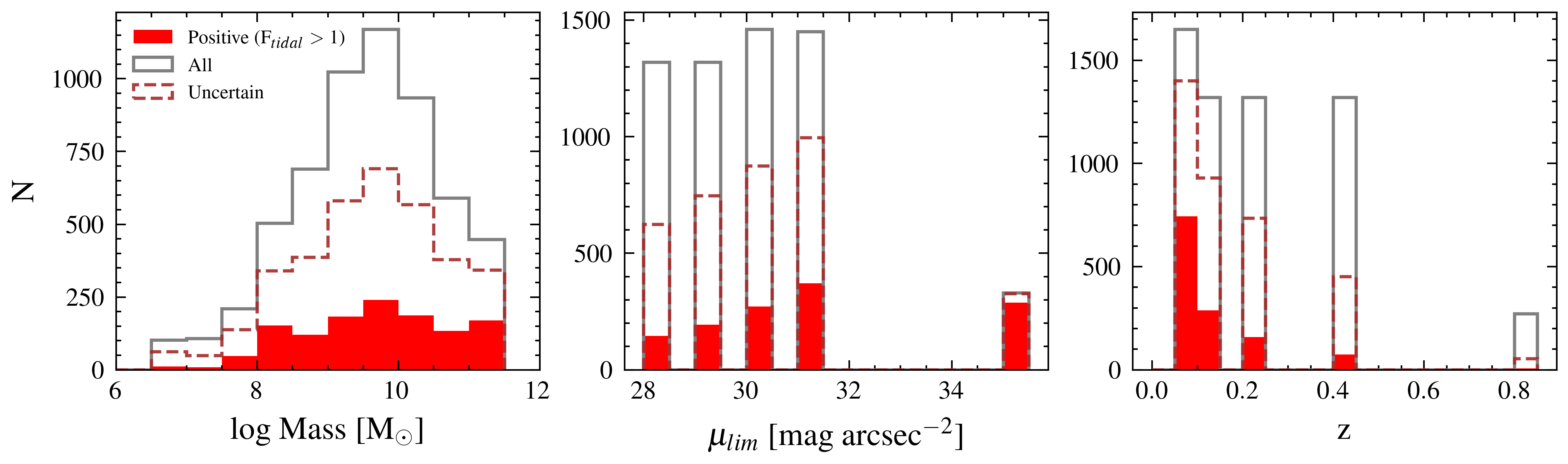}
  \caption{Distribution of stellar mass (left), limiting surface brightness (middle) and redshift (right) of the images from the parent sample presented in \citetalias{Martin2022}. The grey empty histograms represent the full sample (5835 images), the red filled histograms show the images labelled as positive examples of tidal features  (i.e., with  F$_{\rm Tidal}$ > 1), and the brown empty histograms correspond to images with uncertain classifications ( 0 <  F$_{\rm Tidal}$ < 1 or  labelled as \textit{misc/double nucleus} only). Note the large dependence of the fraction of tidal feature identification by the astronomers  with surface brightness limit and redshift (or, equivalently, cutout size).}
  \label{fig:sample}
\end{figure*}

\citetalias{Martin2022} performed a visual inspection of $\sim$ 8\,000\footnote{Note that this number is smaller than 10\,800 quoted on Section \ref{sect:mock-images} due to missing progenitors at some snapshots which are too small to be detected by the structure finder at higher redshift.} unique images by  45 expert classifiers, with a random subset of  them classified six times, by  identifying the presence of tidal features and classifying them into stellar \textit{streams, tidal tails, shells, tidal bridges, merger remnants, double nuclei} or \textit{miscelanea}.\footnote{The \textit{plume} and \textit{asymmetric} categories described in  \citetalias{Martin2022} are combined into a single \textit{miscelanea} category in the catalogue, since there was a large degree of overlap between the two.} The classification was summarized in a  catalogue including 5\,835 unique images. The missing  images were not included in the classification catalogue for being too noisy. We use this catalogue as parent sample in our work. The visually classified images were created at HSC pixel angular scale of 0.2", but rescaled to 1", comparable to the FWHM of the PSF used (an observer might reasonably rebin like this in order to gain additional depth without losing any detail). This rebinning will have a significant impact when applying the models trained with these images to real HSC-SSP data, as we discuss in Section \ref{sect:real-data}. In this work, we use the exact same images as those visually classified to train the deep learning algorithm, so that the labels are consistent.

Figure \ref{fig:labels} shows the result of the classification for a particular galaxy image observed at redshift z=0.05 with different limiting surface brightness. Each image is classified by a number of astronomers (N) which assign the number of observed features (N$_f$) of each class to the image. This means that more than one class can be assigned to each image and that the classification can change with \mulim, but also due to projection effects or spatial resolution (redshift). For this particular example, the deepest image (\mulim=35 mag arcsec$^{-2}$) was classified by N=3 astronomers which annotated features of \textit{streams, shells, tidal tails, merger} and \textit{bridges} adding up to a total of N$_f$=9. On the other hand, the shallower image  (\mulim=28 mag arcsec$^{-2}$) was  classified by N=5 astronomers, of whom only one annotated the feature class of  \textit{merger} (N$_f$=1).

To the best of our knowledge, this is the sample with the largest number of tidal detections visually classified by professional astronomers up to date, making it the optimal sample for training a deep learning algorithm for automated detection of tidal features. However, the  example from Figure \ref{fig:labels} illustrates the challenges of the visual identification of tidal features: the definition of the different classes of features is not objective and there is discrepancy between the classifiers. In some cases, the same images are classified as showing tidal features by some classifiers and as featureless by others. This is a warning about the reliability of the visual classifications and how much they should be trusted as the ground truth. 
Although we are well aware of these caveats, we continue to use this dataset in our analysis as is the largest and most complete galaxy sample with tidal feature labels up to date.

To simplify the problem, in this work we focus on the identification of the presence (or not)  of a tidal feature, regardless of its category and, thus, we consider all the tidal feature classes simultaneously.  Since the images were classified by a varying number of experts, ranging from one to six, and more than one tidal feature category could be assigned to each image, we divide the number of tidal feature identifications by the number of classifiers. We refer to this quantity as the \textit{fraction of tidal detections,} F$_{\rm Tidal}$=N$_f$/N, and consider certain classifications those with  F$_{\rm Tidal}=0$ or F$_{\rm Tidal}$ $\ge $1 (corresponding to 39\% and 38\% of the sample, respectively). For the images with 0 <  F$_{\rm Tidal}$ < 1  (the remaining 22\%) the classification of different experts were inconsistent and we refer to these cases as uncertain classifications. To avoid including uncertain classifications in the loop, we remove the images with  0 <  F$_{\rm Tidal}$ < 1 from the train and test samples. After visual inspection of some images, we found that the classes \textit{misc} and \textit{double nucleus}  do not fit exactly into the tidal features we are aiming to detect. Therefore, images classified \textit{only} as \textit{misc} or \textit{double nucleus}  are also removed from the analysis.

The distribution in mass, surface brightness limit and  redshift of the parent sample and the images classified as showing tidal features are shown in Figure \ref{fig:sample}. The detection of tidal features by humans is largely dependent on the depth of the images and on the image size (or redshift of the galaxy), as reported by \citetalias{Martin2022} and clearly seen in Figure \ref{fig:sample}. This has important consequences for the performance of the algorithm for automated detection of tidal features, as we discuss in Section \ref{sect:results}.

\section{Methodology}\label{sect:cnn}

We use supervised learning for the identification of galaxies showing tidal features; i.e., we need to provide the algorithm with the ground truth we would like to recover in the form of labels (in this case, tidal feature detection or not). We use CNNs, a class of artificial neural networks consisting of convolution kernels that slide along input features and provide feature maps. These maps  are then passed through a fully connected network that outputs a value, corresponding to a particular property that we want to learn. The final function (or weights of the model) is the one that minimizes the difference between the output and the input labels. In this work, the input to the CNN are single-band images in the HSC $r$-band  (we use the $r$-band images since these were the images classified by the professional astronomers in \citetalias{Martin2022}) with their corresponding labels (0s for non-tidal detections and 1s for tidal features). The output of the model, P$_{\rm Tidal}$,  is the probability that the image shows a tidal feature.

\subsection{Image pre-processing}
\label{sect:img}

Before being fed to the CNN, the galaxy images are normalized in the range (0,1)  to avoid operating with very large numbers. For the normalization, the commonly used \textit{asinh stretch} function \footnote{https://docs.astropy.org/en/stable/api/astropy.visualization.AsinhStretch.html} (see \citealt{Lupton2004})  is used, combined with a sigma clipping of  3\% of the faintest and the brightest pixels of each image. This  pre-processing enhances the detection of low surface brightness features. The images are converted to the same size, 69$\times$69 , (the input to the CNN is an array of fixed dimensions)  by rebinning or interpolating the pixel flux, depending on the original image size. We tested  input sizes of 100$\times$100 without obtaining significant changes in the results. Throughout the paper we use the 69$\times$69 stamps as  reference.

\subsection{Input Labels}
\label{sect:labels}

We use a binary classification to separate images which show tidal features (positive samples, labelled as 1s) from images without tidal signatures (negatives, labelled as 0s). Therefore, we unify all classes of tidal features into a single one (detections or non-detections). As explained ins Section \ref{sect:cat-labels}, we use the quantity F$_{\rm Tidal}$ to select positive and negative examples, and leave out of the analysis images with uncertain classifications.

For the  images generated  specifically for this work at \mulim =26, 27 mag arcsec$^{-2}$ , we do not have classifications by the professional astronomers as these images were not part of the original \citetalias{Martin2022} sample. We choose to use the labels of their counterpart images (i.e., with the same galaxy ID, snapshot, redshift and projection) at \mulim=28 mag arcsec$^{-2}$ as their `ground truth'  label.  This exercise allows us to test whether or not the algorithm can recover visually classified  features in images with surface brightness limit 2 mag arcsec$^{-2}$ shallower than the images used for their visual classification.

We randomly split the sample in 85\% for training (resulting in 4\,418 images, out of which 1\,539 are tidal detections) and reserve 15\% for testing (820, out of which 223 are tidals). 

\subsection{CNN architecture}

\begin{table}
\begin{tabular}{lcc}
\hline
 \textbf{Layer Type} & \textbf{Output shape} & \textbf{Parameters} \\
\hline
Input        & (69, 69, 1)   &    0   \\      
\hline
Conv2D       & (69, 69, 32)    &    320      \\  
\hline
MaxPooling2D & (34, 34, 32)   &     0    \\  
\hline     
Conv2D       & (34, 34, 48)   &     13\,872  \\  
\hline   
MaxPooling2D  &  (17, 17, 48)   &     0       \\  
\hline  
 Conv2D       &  (17, 17, 64)    &    12\,352    \\  
\hline 
MaxPooling2D  & ( 8, 8, 64)     &     0      \\  
\hline   
Flatten  &    (4\,096)   &   0       \\  
\hline  
 Dense   &    (64)      &   262\,208    \\  
\hline
 Dense   &    (1)        &         65  \\  
\hline
Total number of parameters &  & 288\,817
\end{tabular}
\caption{Architecture of the CNN used in the main text.}
\label{tab:CNN}
\end{table}

The CNN architecture discussed in the main text, based on the one  presented in \cite{Walmsley2019}, is summarized in Table \ref{tab:CNN}. It consists on three 2D convolution layers with 32, 48 and 64 filters with sizes 3, 3, and 2, respectively and 2$\times$2~\textit{max-pooling} windows. They are followed by a fully connected layer with 64 neurons, Rectified Linear Unit (ReLU)
non-linear activation function and 0.5 dropout rate. A final single neuron outputs values converted to the (0,1) range by applying a sigmoid function. Binary-crossentropy is used as loss function and \texttt{Adam} as optimizer. The number of free parameters of this CNN is 288\,817 (for input sizes 69$\times$69$\times$1).

We have tested other CNN architectures, namely the one commonly used by the authors (e.g., \citealt{DS2018} and \citealt{DS2022}), consisting on four 2D convolutional layers with 32, 64, 128 and 128 filters with sizes 6, 5, 2 and 3, 2$\times$2~\textit{max-pooling} and 0.25 dropout\footnote{This is a slight modification with respect to the original configuration.}, followed by a fully connected layer with 64 neurons and 0.25 dropout. The number of free parameters  is 2\,600\,545, almost ten times larger than in the \cite{Walmsley2019} CNN.  A variation of the \cite{DS2018} architecture, with filter sizes (3, 3, 2, 3) and adding a fully connected layer with 16 neurons before the final layer, has also been tested. In addition, conventional networks such as ResNet-18,-50,-101 \citep{ResNet} and EfficientNet-B0,-B1, -B4 and -B7 \citep{Tan2019} have been attempted. As the use of more complicated CNNs did not significantly improve the results, we have decided to use the \cite{Walmsley2019} architecture as a reference due to its simplicity with respect to other algorithms. 

  We use an overall standard strategy for  training. We train for 100 epochs with a batch size of 100 and a validation split of 0.2 (from the training sample). Data augmentations are performed while training, including vertical and horizontal flip,  weight and height shifts (by 0.05\%),  zoom in and out (0.75-1.3) and rotations (0, 90, 180, 270 degrees). We train 10 independent models, randomly changing the initialization weights and  the training and validation sets. During the training we observed no signs of over-fitting. The results presented in the following sections are based on the average of the output  of the 10 models, which we refer to as P$_{\rm Tidal}$.

\begin{figure*}
  \centering
  \includegraphics[width=0.56\linewidth]{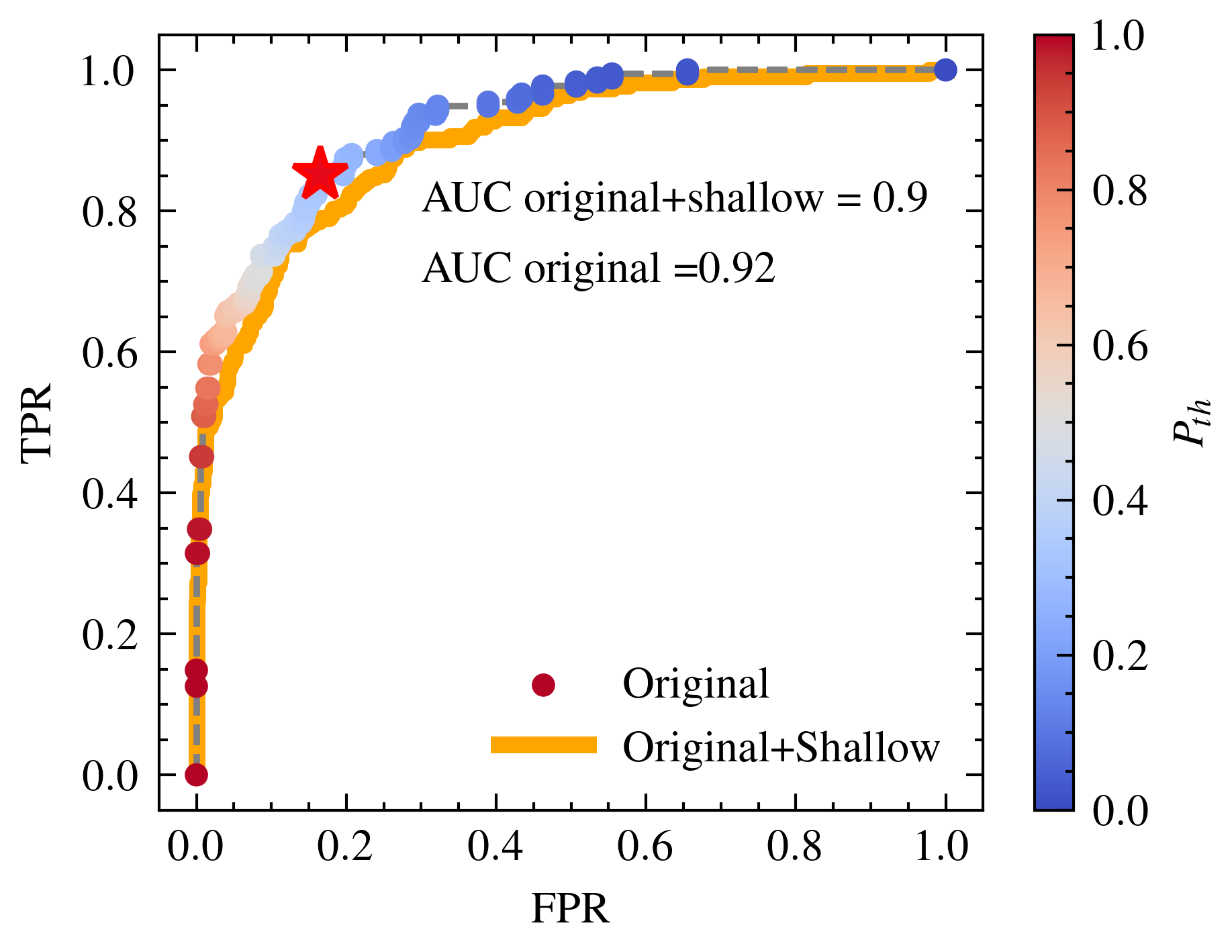}
      \begin{tabular}{c c}
        \textbf{Original} & \textbf{Original+Shallow} \\
          \includegraphics[width=0.45\linewidth]{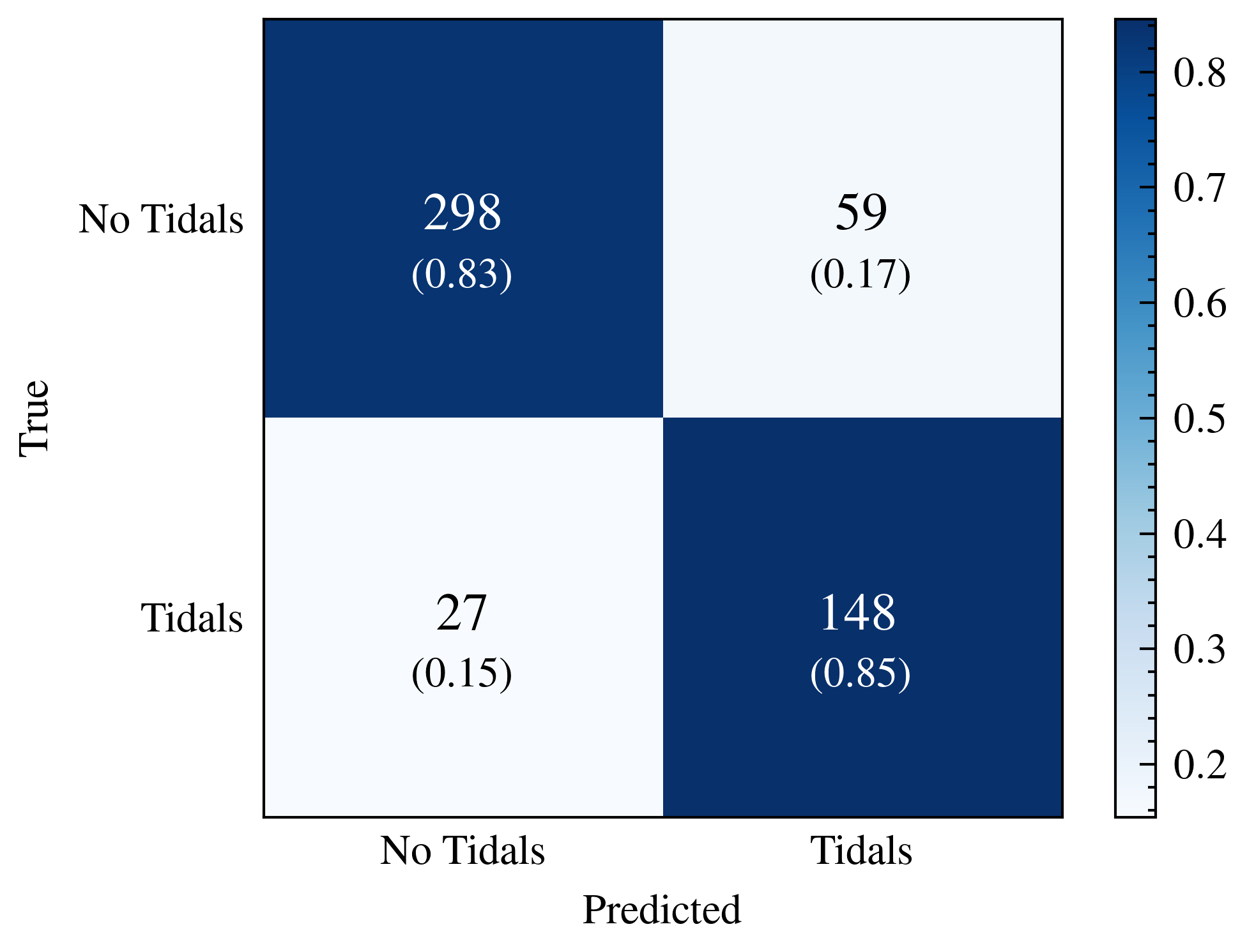} & \includegraphics[width=0.45\linewidth]{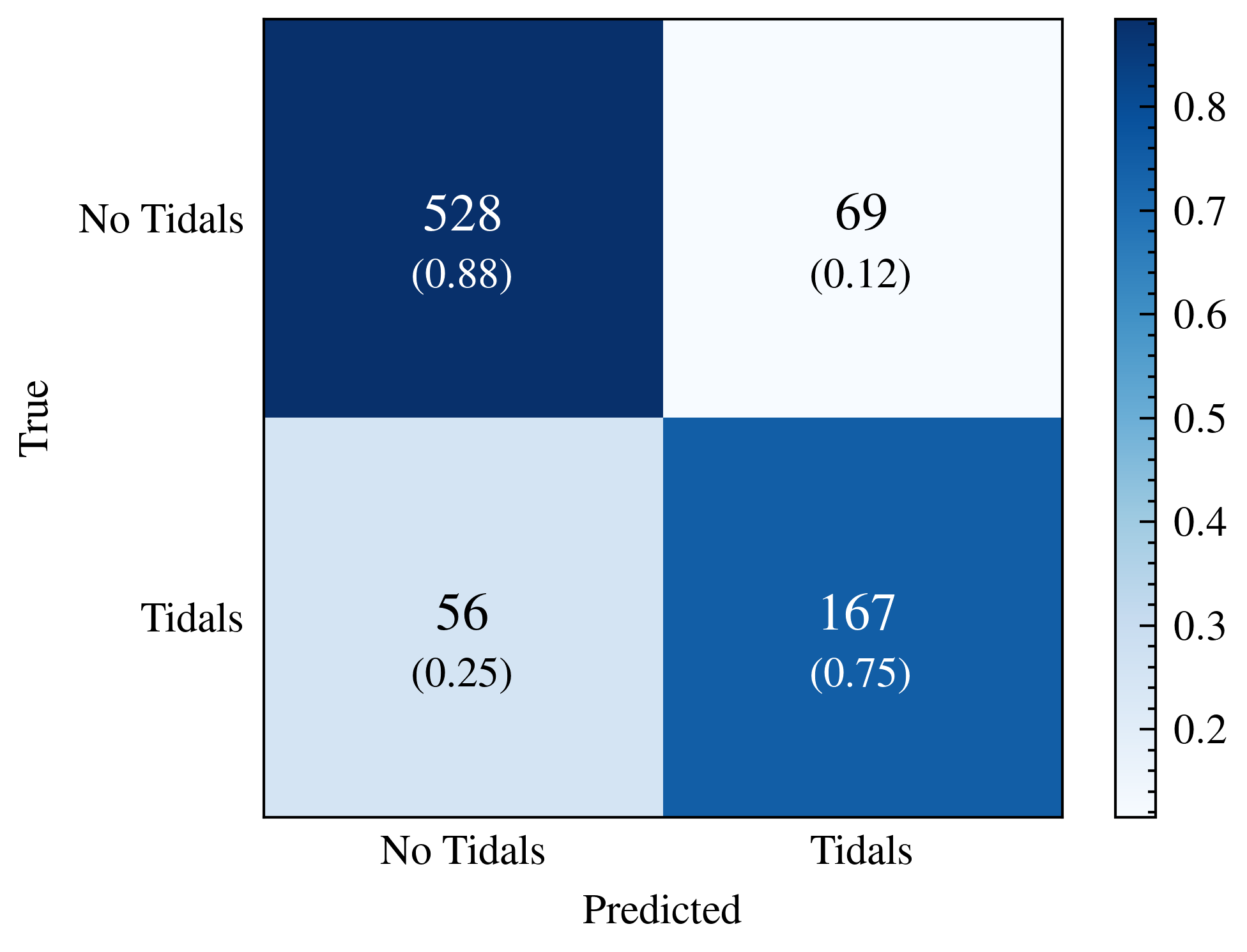}
    \end{tabular}
  \caption{\textit{Upper panel:} ROC curve - True Positive Rate as a function of False Positive Rate - for the \textit{original} (circles, colored coded by P$_{\rm th}$) and \textit{original+shallow} (orange line) test samples. The red star marks the optimal threshold for the \textit{original} sample. \textit{Bottom panels:} Confusion Matrix for the  \textit{original} (left) and \textit{original+shallow} test sample (right) obtained when selecting positive samples as those above the corresponding P$_{\rm th}$ of each sample. Input labels are shown in the $y$-axis, predictions in the $x$-axis. The number of objects is reported in each quadrant, color coded by the fraction of that particular true class (also shown in parenthesis).}
  \label{fig:ROC}
\end{figure*}

\begin{table*}
\begin{tabular}{lccccccc}
\hline
 Test sample & N$_{\rm test}$ & $\%$ Positives & P$_{\rm th}$ & Accuracy & Precision & Recall & F1 \\
\hline
 Original              &  532 & 33  & 0.32 & 0.84  & 0.72  & 0.85 & 0.78  \\
 Original+shallow        &  820 & 27 & 0.31  & 0.85 & 0.71 &  0.75 & 0.73 \\
\hline
\end{tabular}
\caption{Number of galaxies in the \textit{original} and \textit{original+shallow} test samples, and the fraction of those labelled as tidal features, as well as the accuracy, precision, recall and F1 score obtained when selecting as positive predictions the images with model scores above their corresponding P$_{\rm th}$. }
\label{tab:test}
\end{table*}

\section{Results}
\label{sect:results}

\begin{figure*}
  \centering
   \includegraphics[width=0.98\linewidth]{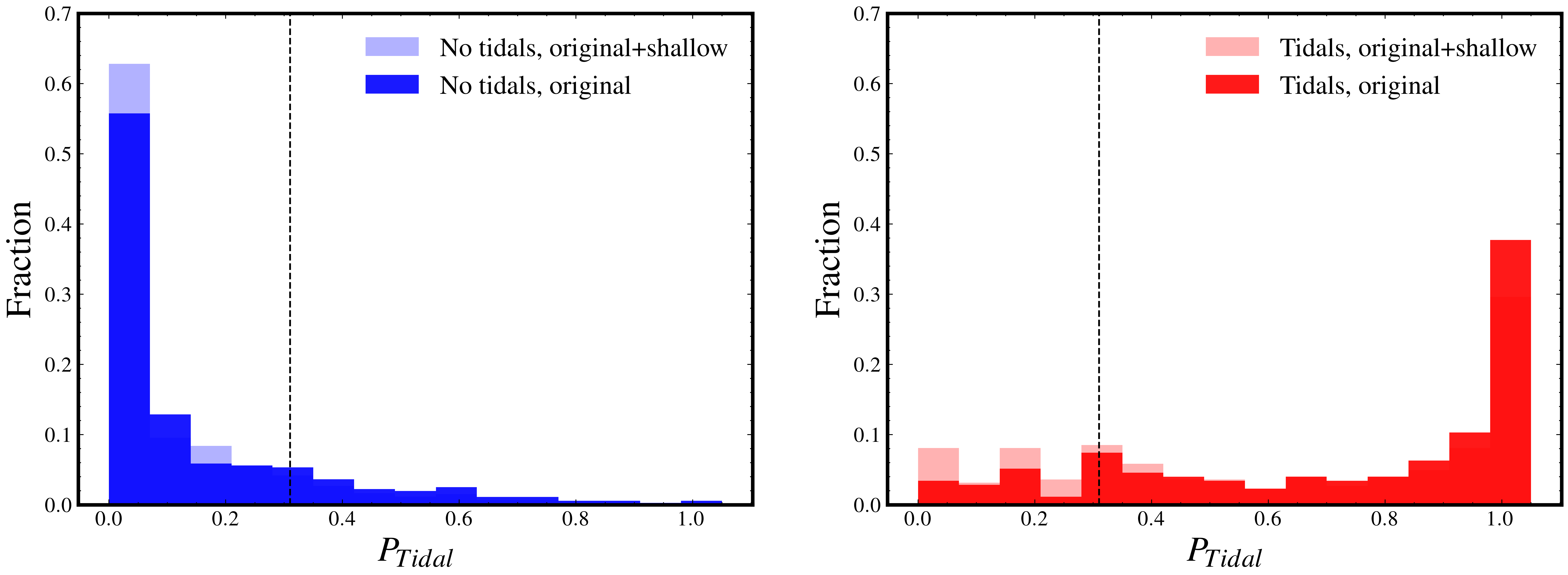} 
    \includegraphics[width=0.97\linewidth]{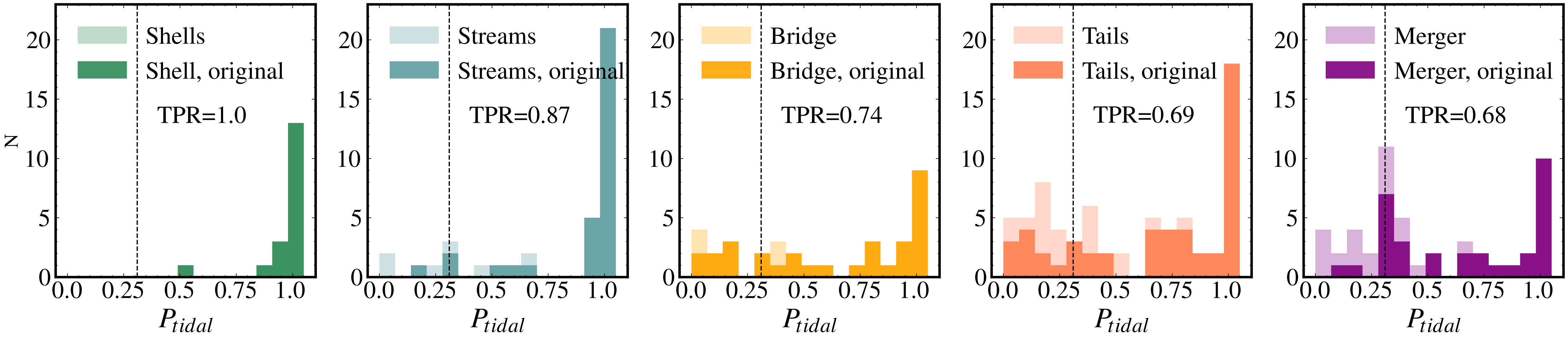} 
  \caption{\textit{Upper panels:} Output probability distribution of the model for tidal detection, P$_{\rm Tidal}$, for the test sample divided into non-tidal visual classifications  (left panel, blue) and tidal visual classifications (right panel, red). Darker colors represent the \textit{original} sample, lighter colors the \textit{shallow} sample. \textit{Lower panels:}  P$_{\rm Tidal}$, for the test sample divided into different categories, as stated in the legend.  The dashed line is P$_{\rm th}=0.31$, the threshold used to define a instance as positive or negative. The true positive rate (TPR) of each category for the \textit{original+shallow} sample is reported in the corresponding panel. }
  \label{fig:class}
\end{figure*}

\begin{figure}
  \centering
  \textbf{True Positive shells}\par\medskip
  \includegraphics[width=0.3\linewidth]{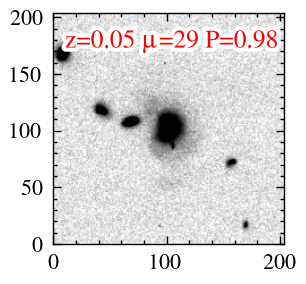}
  \includegraphics[width=0.3\linewidth]{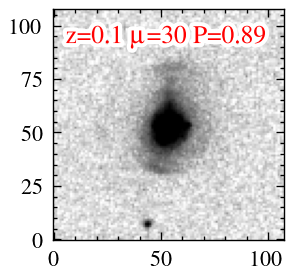}
    \includegraphics[width=0.3\linewidth]{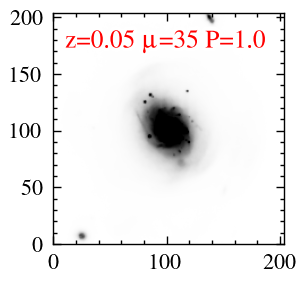}
    
  \textbf{True Positive streams}\par\medskip
  \includegraphics[width=0.3\linewidth]{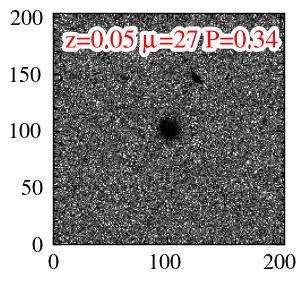}
 \includegraphics[width=0.3\linewidth]{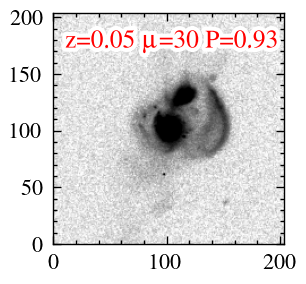}
  \includegraphics[width=0.3\linewidth]{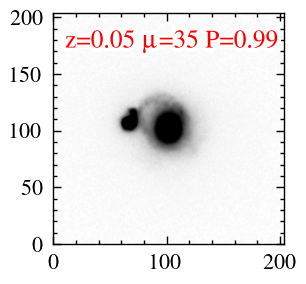}

  \textbf{True Positive mergers}\par\medskip
\includegraphics[width=0.3\linewidth]{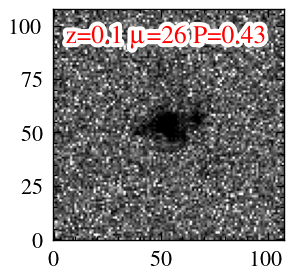}
   \includegraphics[width=0.3\linewidth]{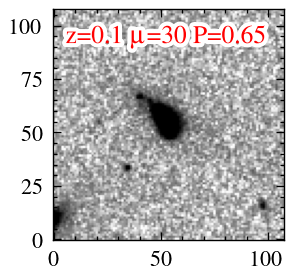} 
  \includegraphics[width=0.3\linewidth]{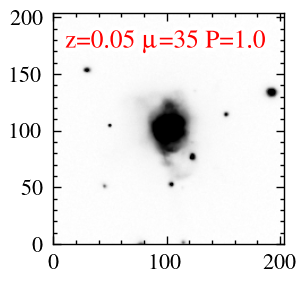}

  \caption{TP examples of shells, streams and mergers, from top to bottom. The cutouts have been processed as described in Section \ref{sect:img}, but are shown at their original sizes (i.e., they are not binned to 69$\times$69, which is the input to the CNN), and the $x$ and $y$ axes correspond to the number of pixels. The information shown in each cutout corresponds to the redshift, the  surface brightness limit (increasing from left to right) and the output probability of the model, P$_{\rm Tidal}$.}
  \label{fig:examples_TP}
\end{figure}

\begin{figure}
  \centering

\textbf{False Negative streams}\par\medskip
  \includegraphics[width=0.3\linewidth]{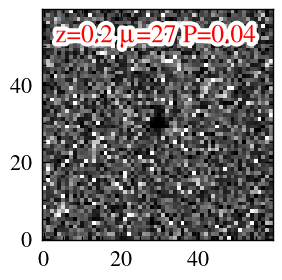}
   \includegraphics[width=0.3\linewidth]{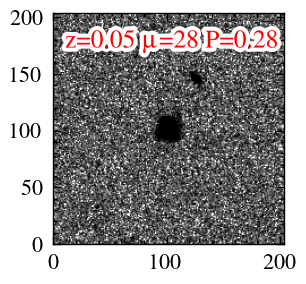}
    \includegraphics[width=0.3\linewidth]{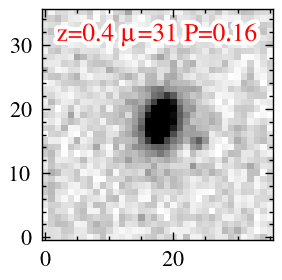}
    
\textbf{False Negative mergers}\par\medskip
  \includegraphics[width=0.3\linewidth]{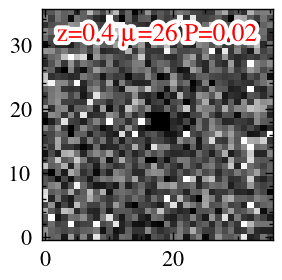}
  \includegraphics[width=0.3\linewidth]{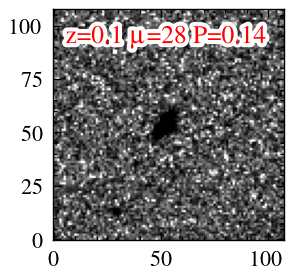}
   \includegraphics[width=0.3\linewidth]{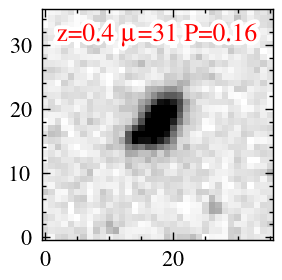}
  
  \caption{Same as figure \ref{fig:examples_TP} but for  FN examples of \textit{streams} and \textit{mergers} (top and bottom panels, respectively). Note that there are no  false negative shells (see Figure \ref{fig:class}). The features are hard to detect by eye, so it is not surprising that the model fails in these cases.}
  \label{fig:examples_FN}
\end{figure}

\begin{figure}
    \includegraphics[width=0.9\linewidth]{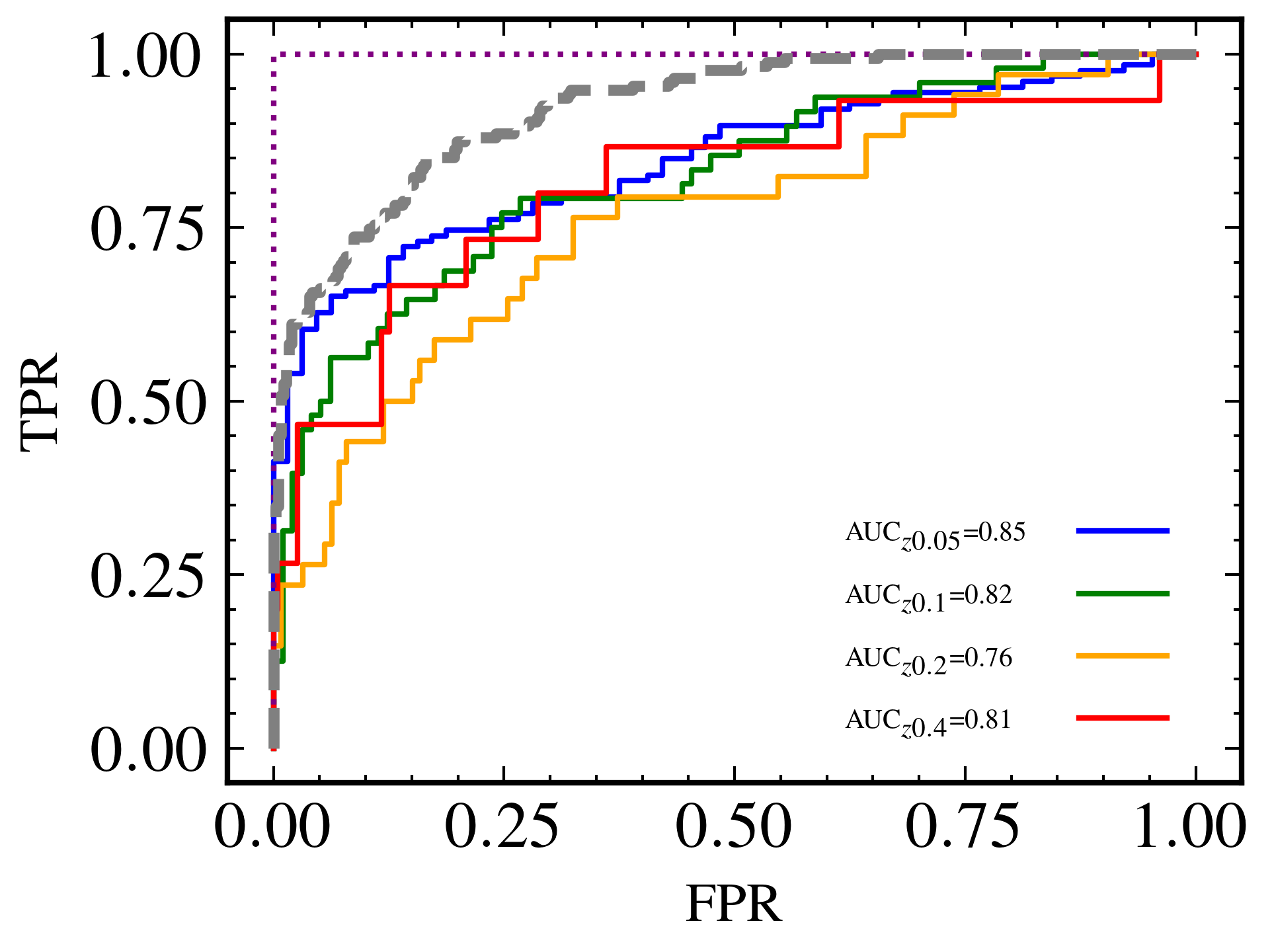}\par 
\caption{ROC curves for galaxies at different redshifts, color coded as indicated by the legend. The gray dashed line shows the ROC curve for the full \textit{original} sample. Note that there are no positive cases at $z$=0.8, and the model correctly classifies all the images at this $z$ as negatives; hence we represent the ROC curve as the dotted purple line.} \label{fig:ROC-z}
\end{figure}

\begin{figure}
    \includegraphics[width=0.9\linewidth]{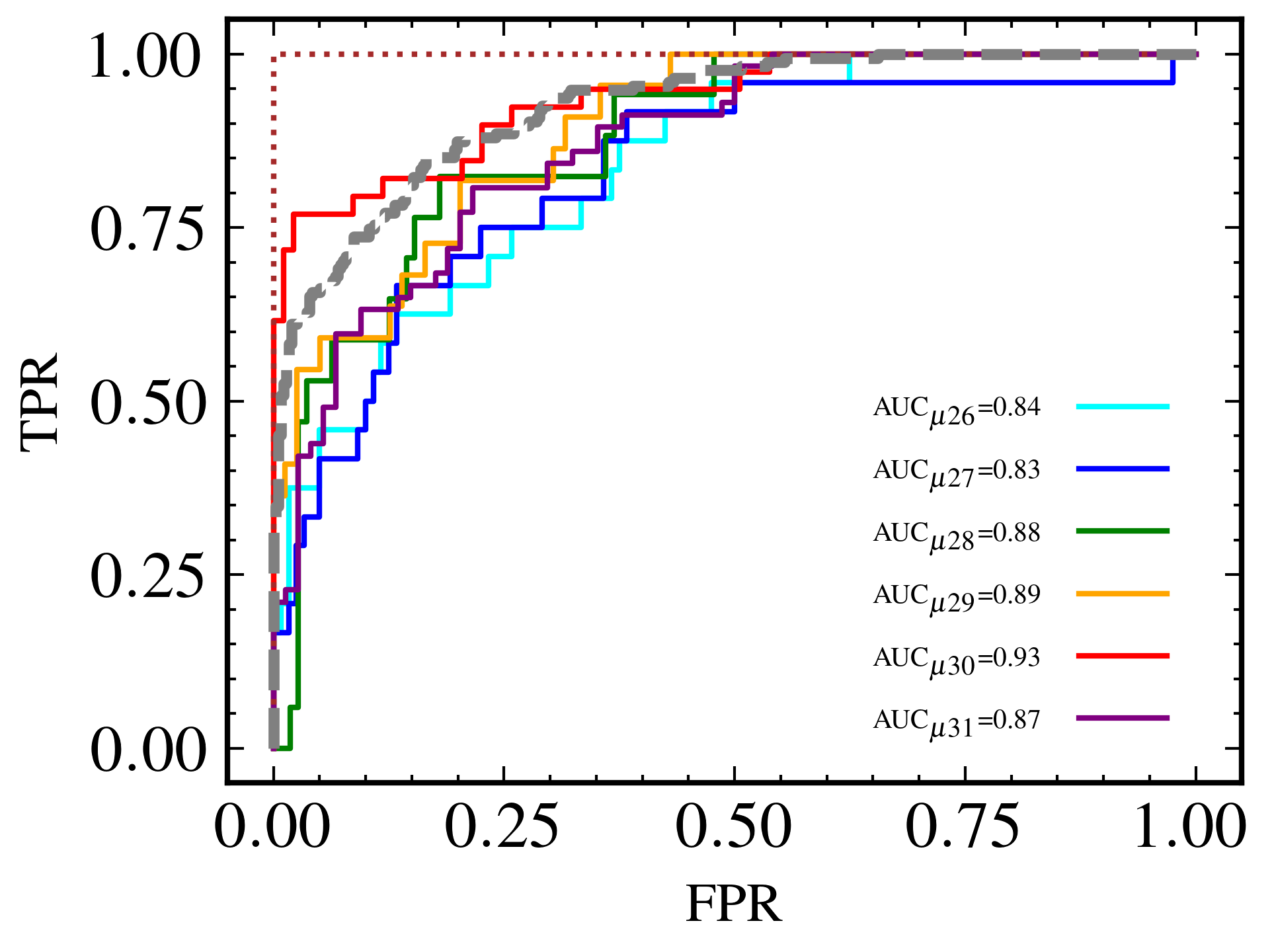}\par 
\caption{Same as figure \ref{fig:ROC-z}  for images at different surface brightness, color coded according to the legend. Note that at $\mu$=35 mag arcsec$^{-2}$ all images show tidal features and are correctly classified as such by our model (we plot the ROC curve as the brown dotted line)}. \label{fig:ROC-SB}
\end{figure}

In this section, we study the performance of our models when applied to the test dataset. We consider two different tests sets: the one containing only the original simulations and labels by professional astronomers (i.e., surface brightness \mulim $\ge$ 28 mag arcsec$^{-2}$) and the test set which includes the original and the simulated images at \mulim=26, 27 mag arcsec$^{-2}$. We will refer to the former as the \textit{original} test sample and to the latter as the \textit{original+shallow} test sample.

We use standard metrics for studying the performance of the models:\\

\begin{equation}
   \rm Accuracy=\frac{TP+TN}{TP+TN+FP+FN}=\frac{TP+TN}{Total}\\
\label{Eq:Acc}   
\end{equation}

\begin{equation}
   \rm  Precision=\frac{TP}{(TP+FP)}=\frac{TP}{P_{pred.}}\\
\label{Eq:P}      
\end{equation}

\begin{equation}
   \rm  Recall=\frac{TP}{(TP+FN)}=\frac{TP}{P_{input}}\\
\label{Eq:R}      
\end{equation}

\begin{equation}
   \rm  F1=2\times\frac{P\times R}{(P+R)}\\
\label{Eq:F1}      
\end{equation}

where TP, TN, FP and FN stand for true positives,  true negatives, false positive and false negative, respectively, while P$_{\rm pred}$ and P$_{\rm input}$ are the total number of predicted and input positives, respectively. To separate the instances into positive and negative predictions, we use the \textit{binary classification threshold probability}, P$_{\rm th}$, which is the value that optimizes the true positive rate (TPR, i.e.  the fraction of correctly identified  tidal detections) and the false positive rate (FPR, i.e., the fraction of non tidals classified as tidal detections) simultaneously. The number of galaxies in each test sample and the fraction of positives (labelled as tidal detections in the input catalog), as well as the accuracy, precision, recall and F1 score of each sample is reported in Table \ref{tab:test}. The accuracy is the fraction of correctly classified instances, the precision is the fraction of TP among the instances classified as positive (analogue to the purity), while the recall is the fraction of TP among the positive input instances (analogue to the completeness). Finally, the F1 score is the harmonic mean of the two.

Figure \ref{fig:ROC} shows the receiver operating characteristic curve (ROC) that represents TPR  versus FPR  as the discrimination threshold (P$_{\rm th}$) is varied. An adequate classifier  would maximize the TPR while keeping the FPR low. The area under the ROC curve (AUC) is above 0.9 in both cases (a perfect classifier would have AUC=1). We also show the confusion matrices for the two test samples using as probability threshold the optimal value for each  sample  to separate the predictions into positive and negative classes. As reported in table \ref{tab:test},  the accuracy (Equation \ref{Eq:Acc})  for the \textit{original} and  \textit{original+shallow} test samples is 0.84 and 0.85, while the precision (or purity, Equation \ref{Eq:P}) is 0.72 and 0.71, respectively. These values are surprisingly similar, taking into account the inclusion of $\sim$ 300 images with \mulim < 28 in the \textit{original+shallow} test sample for which the ground truth is assumed to be the labels at  $\mu_{lim}$ = 28 , i.e., those reported for images two magnitude deeper than the actually classified images. The main difference is in the recall (or completeness, Equation \ref{Eq:R}), that drops from 0.85  for the \textit{original} test sample to and 0.75 for the \textit{original+shallow} test sample. As expected, it is more difficult for the algorithm to recover tidal detections  in shallower images. We discuss  the surface brightness dependence of the classification in Section \ref{sect:SB}. Since the precision values are very similar for the two test samples but recall is smaller for the \textit{original+shallow} test sample, the F1 score (Equation \ref{Eq:F1}) is also lower for the \textit{original+shallow} (F1=0.73) than for the \textit{original} sample (F1=0.78).

\subsection{Dependence on tidal feature class}
\label{sect:class}

Now we study the ability of our CNN to detect different classes of tidal features. Figure \ref{fig:class} shows the output probability of our model, P$_{\rm tidal}$ (larger values correspond to more confident  detection of tidal features), divided in the classes provided in the \citetalias{Martin2022} catalogue. Clearly, there are classes which are easier to identify  than others. For example, all the \textit{shells} in the test sample are recovered,  and the  TPR for the \textit{streams} is 0.84, while for \textit{mergers} or \textit{tidal tails} is below 0.7. It is also evident from Figure \ref{fig:class} that the model performs worse for the \textit{shallow} sample: the P$_{\rm tidal}$ values are lower for the positive cases of this sub-sample. We discuss in more detail the effect of the \mulim~in section \ref{sect:SB}.

 Figure \ref{fig:examples_TP} shows representative examples of TP identifications of \textit{shells, streams} and \textit{mergers} at different surface brightness limits. The features are very evident for the deep images (right panels) and, in some cases, it is surprising that the model is able to identify the tidal features in the more noisy images (left panels). Besides, the cutouts are displayed at their original size, not binned to 69$\times$69, which is the input to the CNN. These examples show that \textit{shells} and \textit{streams} are easier to identify by eye than \textit{mergers\textit}. We note, however, that the number of \textit{shells} in the test sample is small (12) and that there are no \textit{shells} in the \textit{shallow} test sample. This is due to the fact that there are no visually identified \textit{shells} in images shallower than \mulim=28 mag arcsec$^{-2}$, from which the labels for the \textit{shallow} sample come from. In other words, the fact that we are recovering more \textit{shells} may be due to these structures being identified by eye only in the deeper images. We note again  that the different classes reported in the \citetalias{Martin2022} catalogue are not mutually exclusive (see also Figure \ref{fig:labels}), and therefore images identified as \textit{shells} can also fall into other categories (this happens indeed for 10 out of 12  \textit{shells} in the test sample).

FN cases are shown in Figure \ref{fig:examples_FN} for \textit{streams} and \textit{mergers}. As  all the \textit{shells} in the test sample are correctly identified by our model, there are no FN for this category. Even for the deeper images (right panels) it is difficult to identify the tidal features, while the shallower images (left panels) are dominated by noise. Therefore, it is
not surprising that the model fails in these cases.

\subsection{Dependence with redshift}
\label{sect:z}

In this section, we report the performance of the model when the test sample is divided into different redshift bins, which is directly related to the original stamp size (the stamps are then resampled into 69$\times$69 because the input to the CNN has fixed dimensions, see details in Section \ref{sect:data} and \ref{sect:img}).  Figure \ref{fig:ROC-z} shows the ROC curves as a function of redshift. As expected, the larger AUC is obtained in the lower redshift bin ($z$=0.05). However, the dependence on redshift is not very strong and not even linear. For example, the AUC is larger for $z$=0.4 (AUC=0.81) than for $z$=0.2 (AUC=0.76), probably due to the lower fraction of visual detections at higher redshifts, which increases the overall accuracy. This is evident for $z$=0.8 (shown as a dotted line), where there are no galaxies classified as tidal detections in the test sample, neither in the input labels nor by our model, and therefore the accuracy is 100\%.

\subsection{Dependence with surface brightness}
\label{sect:SB}

Finally, we study the dependence of the model performance on the surface brightness of the images to be classified. Figure \ref{fig:ROC-SB} shows the ROC curve for the different \mulim values. The lower AUC correspond to the shallower images ($\mu$=26, 27), as expected, and the best results are obtained for $\mu$=30 mag arcsec$^{-2}$.  We  show the ROC curve for  $\mu$=35 mag arcsec$^{-2}$ as dotted line because all the images from the test sample are classified as tidal detections, both in the input catalogue and by the model (just the opposite of what happens at $z$=0.8). Surprisingly, the AUC at  $\mu$=31 mag arcsec$^{-2}$ is lower (AUC=0.87) than at  $\mu$=30 mag arcsec$^{-2}$ (AUC=0.93) and comparable to the values obtained at  $\mu$=28 mag arcsec$^{-2}$. 

\begin{figure}
\centering

    $\mu=26$ mag arcsec$^{-2}$\par\medskip
    \includegraphics[width=0.8\linewidth]{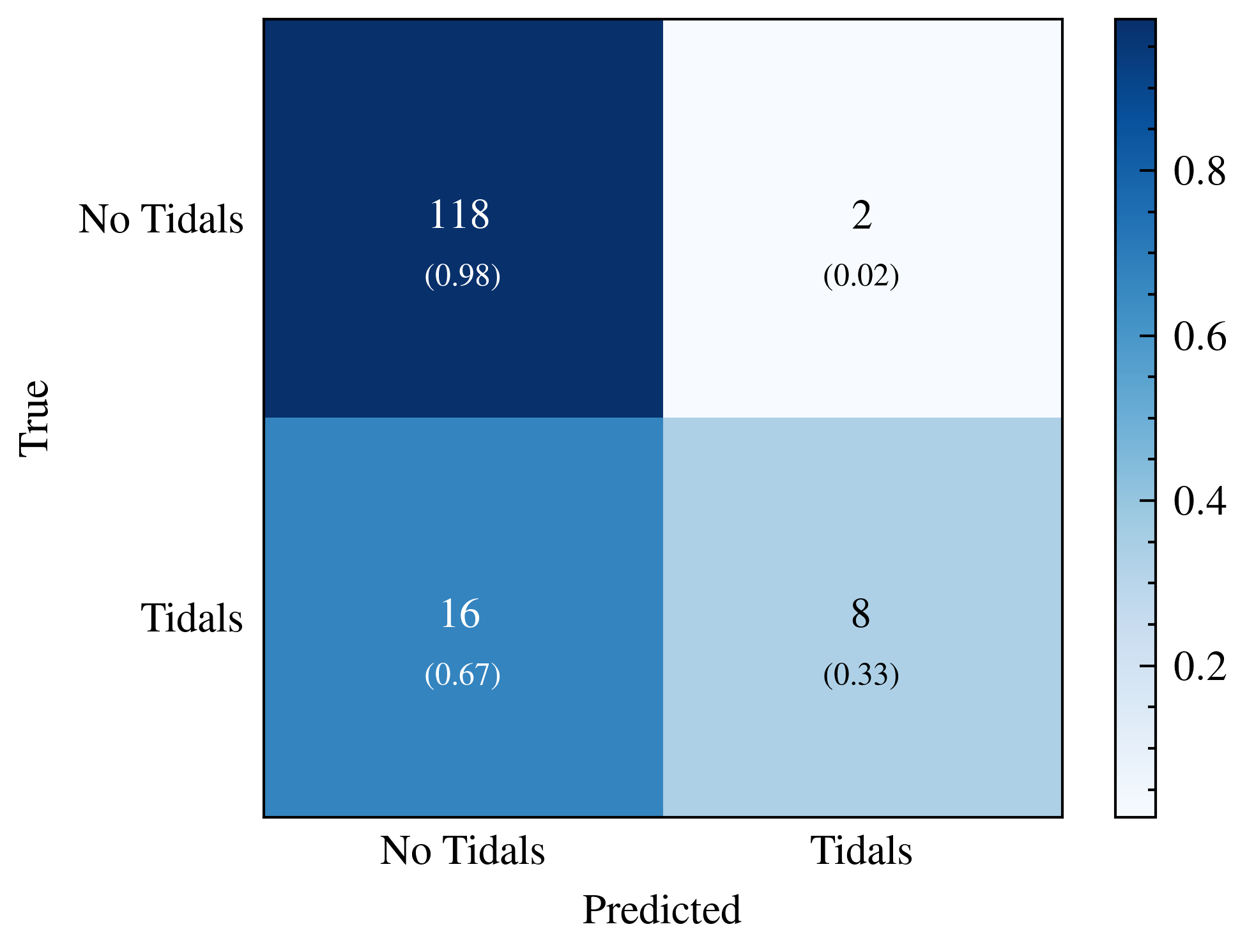}\par

$\mu=30$ mag arcsec$^{-2}$\par\medskip
    \includegraphics[width=0.8\linewidth]{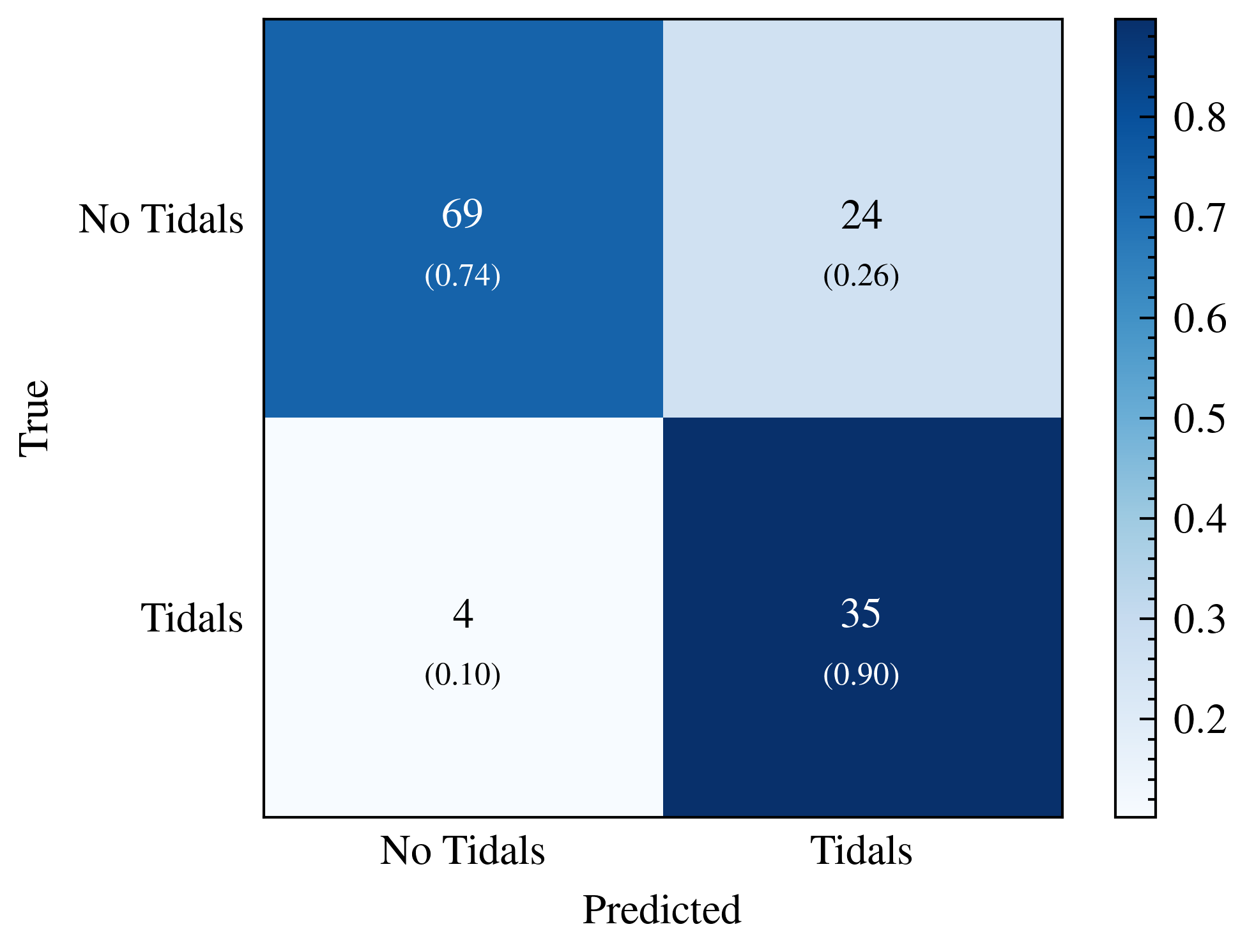}\par

$\mu=31$ mag arcsec$^{-2}$\par\medskip
    \includegraphics[width=0.8\linewidth]{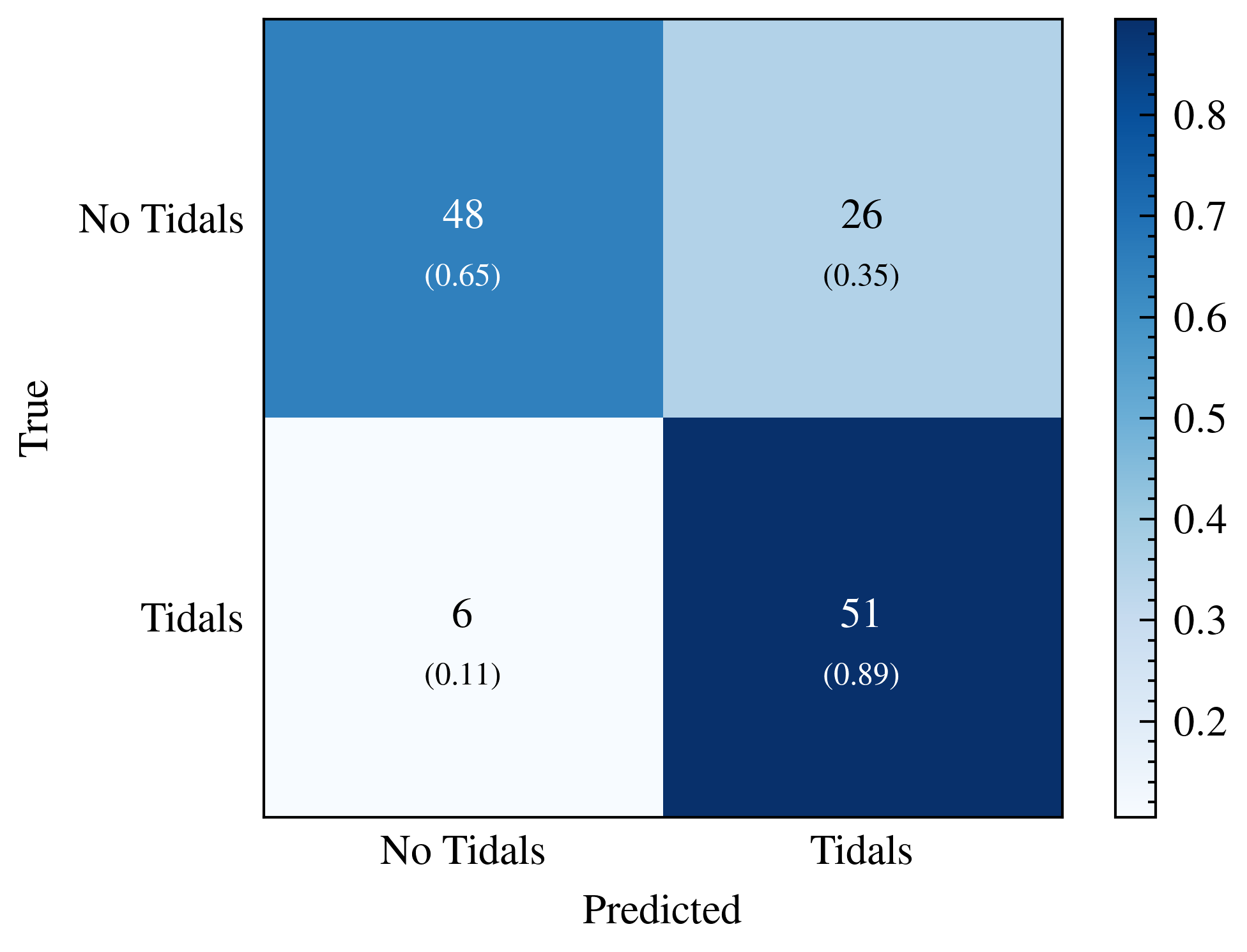}\par

\caption{Confusion matrices for three different surface brightness bins: \mulim=26, 30, 31 mag arcsec$^{-2}$, from top to bottom. The number of objects is reported in each quadrant, color coded by the
fraction of that particular true class (also shown in parenthesis).} 
\label{fig:CM-SB}
\end{figure}

To shed more light on the classification efficiency, Figure \ref{fig:CM-SB} shows the confusion matrices (generated by setting P$_{\rm th}$=0.31) for three surface brightness limits ($\mu$=26, 30, 31 mag arcsec$^{-2}$). These confusion matrices highlight the fact that the large AUC for  $\mu$=26 mag arcsec$^{-2}$ is mainly driven by the ability of the classifier to correctly identify images without tidal detections (98\% accuracy for the negative subsample), while it struggles to correctly classify the tidal detections (only 33\% are recovered in this surface brightness range). On the other hand, at    $\mu$=31 mag arcsec$^{-2}$ the contrary happens: the model is able to correctly identify 89\% of the tidal detections, at the cost of misclasifying 35\% of the non-detections. At $\mu$=30 mag arcsec$^{-2}$ the model is able to correctly identify 90\% of the tidal detections while keeping the contamination (i.e., the number of FP) at 26\%. 

\begin{figure}
    \includegraphics[width=0.9\linewidth]{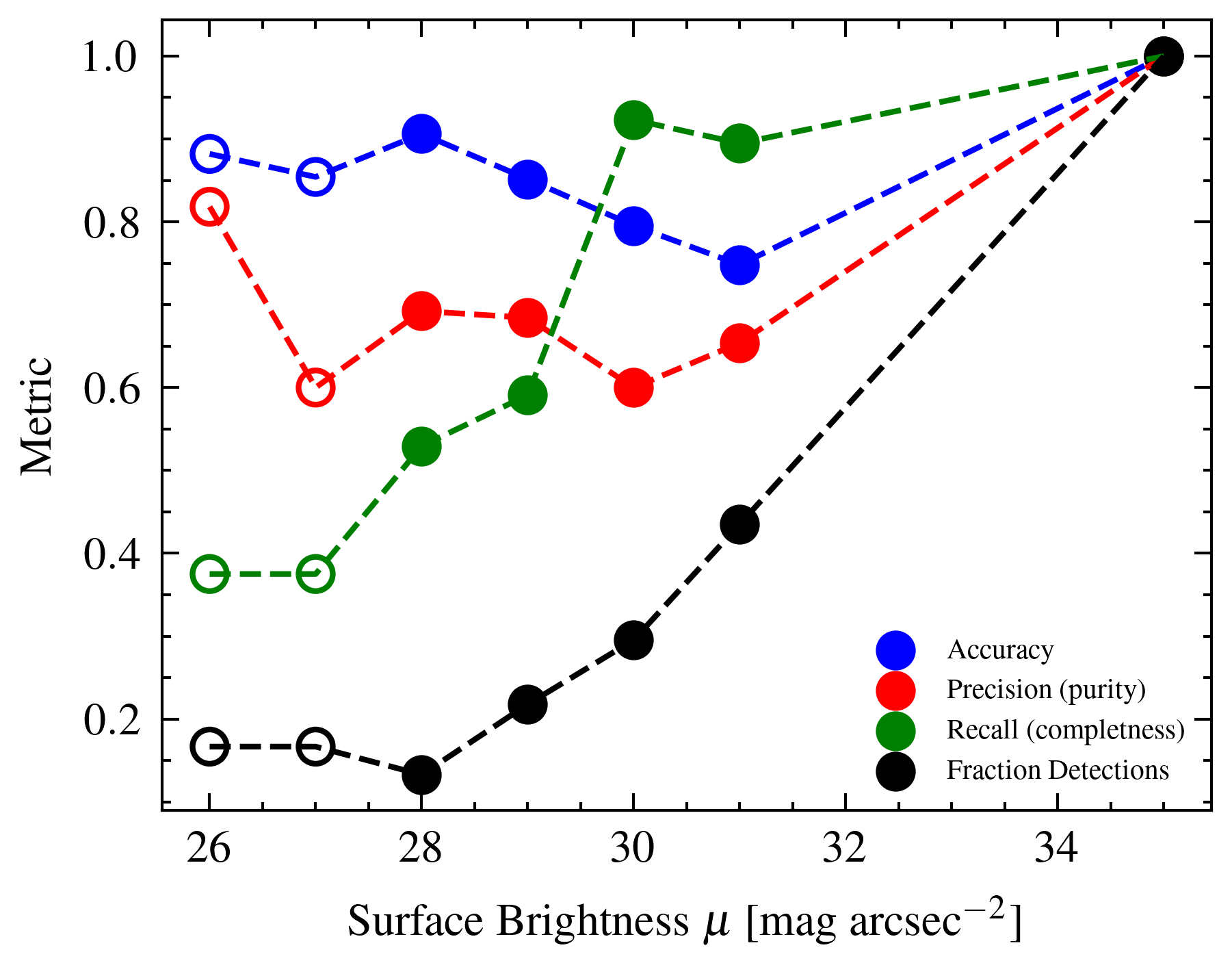}\par 
\caption{Accuracy (blue), precision (red) and recall (green) obtained for the test sample as a function of the images' surface brightness \mulim. The fraction of tidal detections in the input catalogue is shown in black. The bins at \mulim=26, 27 mag arcsec$^{-2}$ are plotted as empty circles to highlight the fact that they are not part of the original sample from \citetalias{Martin2022}; the labels used for training and testing are the ones for their corresponding images at $\mu$=28  mag arcsec$^{-2}$ .} \label{fig:Acc-SB}
\end{figure}

While this trend could be expected, and is in line with the larger fraction of tidal detections obtained in deeper images by visual inspection (see Figure \ref{fig:sample}), it could be an indication that our model has, at least to some extent, learned the signal-to-noise of the images: it tends to classify  deeper images  more frequently as tidal detections. To test this assumption, Figure \ref{fig:Acc-SB} shows the accuracy, precision and recall for each surface brightness bin, as well as the fraction of positive samples  (tidal detections) in the input catalogue.

As reflected by the confusion matrices, the recall (completness) of the model is highly dependent on the depth of the images to be classified, going from around R~$\sim$~0.40 at $\mu$=26 mag arcsec$^{-2}$ to R~$\sim$ ~0.90 at  $\mu$=31 mag arcsec$^{-2}$. However, this is not as clearly reflected in the precision (purity) values,  roughly constant and above  P~$\sim$~0.60 at all surface brightness. In the same way, the accuracy is stable throughout the whole magnitude range ($\sim$80\%), indicating that the model has not simply learned the signal-to-noise of the images.  The fact that the accuracy does not improve with  \mulim, even if the recall (completness) does, can  be explained by a larger fraction  of FP. However, the increase in FP does not decrease the precission (purity) because of the larger fraction of  tidal detections in the input catalogue at higher \mulim  (black symbols in Figure \ref{fig:Acc-SB}), which together with the larger recall, increase the number of TP to counterbalance the larger number of FP. 

We would like to highlight here again that the labels used for the \textit{shallow} sample (\mulim <28 mag arcsec$^{-2}$) correspond to the visual classification of the images with $\mu$=28 mag arcsec$^{-2}$, which explains the drop in completeness in this surface brightness regime. Indeed, it is remarkable that our model is able to recover 40\% of the images with tidal detections, even when the images are 2 orders of magnitude shallower that the ones used for labelling. This is in agreement with recent results suggesting that CNNs trained with `intrinsic' ground truth can recover astronomical features hidden to the human eye (see e.g., \citealt{VegaFerrero2021}). This result could have a large impact on  the design strategy of future surveys.

We emphasize that redshift and surface brightness limits are  intertwined in the current analysis. Unfortunately, the test sample size is too small to examine the trends at each \mulim 
 limit separately,  at fixed redshift (or viceversa).

\section{Application to real HSC-SSP data}
\label{sect:real-data}

Our models are trained in HSC-like mock images. Therefore,  it is important to test the performance of the algorithm in real data with similar characteristics to the training dataset. We thus use images from the HSC-SSP survey (Hyper Suprime-Cam Subaru Strategic Survey) Wide layer \citealp{Aihara2018, Aihara2018b}. The
HSC Wide layer covers the largest on-sky area at a relatively
shallow depth ($i\sim$26) relative to the Deep and UltraDeep
Layers ($i\sim$27 and $i\sim$28, respectively).

In particular, we have applied our models to the HSC-SSP images presented in \cite{KadoFong2018}. These are $\sim$21,000 galaxies from the internal data release S16A, with spectroscopic redshifts from SDSS at 0.05 $< z < 0.45$. \cite{KadoFong2018} used a filtering algorithm to identify tidal features, combined with visual classification of the  features detected by such filtering, resulting in a sample of $\sim$1,200 tidal feature detections. Our first approach to this work was to use the labels from \cite{KadoFong2018} sample to train the algorithm for tidal stream identification, but after exhaustive testing, the results were not good enough (F1=0.37), mostly due to the small completeness achieved by the models (R=0.26). 

As previously explored in the context of tidal feature classification of mock images (Section \ref{sect:cat-labels}), one source of significant label noise is the reliability of visual inspection itself. Visual inspection of the full sample by three professional astronomers (I.D., H.S., O.K.L.) provided a new classification, which revealed that only 1/4 of the galaxies showing tidal features according to the filtering algorithm were classified as such by all visual classifiers. We believe that the combination of noisy labels, small positive training sample and low surface brightness features to be detected limited the performance of the algorithm. This was the main reason why we decided to use the HSC-like mock images for training the CNN.  We note that the original classification by \cite{KadoFong2018} used both the images and the output of the filtering algorithm in order to detect features close to the host galaxy, so it is expected that the re-classification of the images alone would yield a lower tidal feature detection rate. On the other hand, as the original \cite{KadoFong2018} sample focused on the comparison of the properties of stream and shell hosts, the visual inspection was performed only for the images where a tidal detection  was identified by their filtering algorithm, meaning that the completeness of the sample is uncertain.

It is well known that deep learning models are sensitive to the characteristics of the training sample and that thus techniques such as transfer learning (e.g., \citealt{DS2019a})  should be used for optimizing models trained on different datasets than the target sample. We attempted, without success, to use transfer learning to fine tune the models on real HSC-data, making exhaustive tests on the number of layers to be trained, the learning rate, etc. The best result we obtained is AUC=0.64, and the output values are concentrated towards the lower end (P$_{\rm Tidal} < 0.7$ ). These poor results emphasize the strong dependence of the model performance on the data they are trained with and warns us against carelessly applying models to different data domains. It also suggest that mock images from simulations are not as realistic as we might have expected.

A possible explanation for the bad performance could be the differences in the way the mock images were created compared to real HSC data. For example, the angular resolution of the HSC-mock images is poorer (1" vs 0."167 for real HSC-SSP). In addition, the simulated images do not include real backgrounds, processing artifacts, contaminating sources, or sky subtraction residuals. This may compromise the model's ability to assess real data, as already discussed in \cite{Bottrell2019}. The morphological classification of TNG simulated images presented in \cite{HuertasCompany2019} was significantly improved by adding realism to the simulated images. Unfortunately, adding realism to the images could change the visual classification used as `ground truth': some features could become undetectable in the presence of brighter objects. Therefore, training the CNN with more realistic mock images and new labels is beyond the scope of the current analysis. We will investigate these possible improvements in the forthcoming studies. 

The reason why transfer learning might not solve the domain shift could be the fact that the \mulim in the simulated images is not fixed. This implies that the model needs to transfer from many sparsely sampled domains to another, instead of transferring from one well-sampled domain to another. Intuitively, the former may be harder. Another aspect which could  have a significant effect is the small parent sample of galaxies used to produce the simulations, based on 36 galaxies only that may not represent the diversity of real galaxy populations.  Since the observed sample is flux limited and the simulated sample is volume limited, the former should have a flatter mass distribution with more massive galaxies. Also, the real data have continuous redshift distribution, while the mock images are simulated in five redshift bins.
All these differences combine together and it is not possible to investigate each aspect separately with the current sample. Thus, we cannot pinpoint the property that is the main cause of the poor model performance across domains.

\section{Summary and conclusive remarks}
\label{sect:summary}

Tidal interactions are expected to play a critical role in galaxy mass assembly and evolution, but their  low surface brightness  make these features difficult to detect. Automated methods for the identification and classifications of tidal features will be compulsory for the analysis of large upcoming surveys such as LSST or Euclid. 

In this work, we take advantage of the catalogue presented in \citetalias{Martin2022}, that provides tidal feature classifications by professional astronomers for a sample of $\sim$6000 galaxy images from the  \textsc{Newhorizon} simulations. This constitutes the largest catalogue of visual identifications of tidal features up to date. The galaxies are simulated at different evolutionary times and redshifts and HSC-like mock images with different surface brightness limits (\mulim=28-35 mag arcsec$^{-2}$) were visually inspected by a varying number of  professional astronomers (ranging from 2 to 6). 

We use a CNN to train a supervised deep learning binary model which aims to reproduce human visual identification of galaxies with tidal features. For this, we have labelled as positives all the images for which  a tidal feature was identified by all the classifiers, regardless of the tidal feature category (F$_{\rm tidal}$ > 1, as detailed in Section  \ref{sect:labels}) and as negative those with no tidal identification (F$_{\rm tidal}$ = 0). We do not use galaxies for which the presence of a tidal feature was uncertain (disagreement between classifiers). In addition to the original sample, we have created shallower images, more similar to current available observations, at \mulim=26, 27 mag arcsec$^{-2}$. These images were not classified in
\citetalias{Martin2022} and we use their corresponding labels at \mulim=28 mag arcsec$^{-2}$ as ground truth. We remark the fact that, since the visual classifications are used as input label to the model, any bias present in human classification would be passed on to the deep learning algorithm (see, for example, how the fraction of tidal detections depends on the image properties in Figure \ref{fig:sample}). Our main conclusions are:

\begin{itemize}
    \item The deep learning model is successful in reproducing the human identification of images with tidal features in the HSC-mock images, reaching accuracy, precision and recall values of Acc=0.84, P=0.72 and R=0.85 for the original test sample, using the optimal threshold, P$_{\rm th}$=0.31, to select positive cases of tidals.
    \item The results are surprisingly similar in terms of global accuracy and purity (Acc=0.85, P=0.71) when the shallower test sample is included, even though these numbers are computed corresponding to the labeling of images one or two orders of magnitude deeper. 
    \item The completeness of the model for the \textit{original+shallow} test sample is smaller than for the \textit{original} one (R=0.75 vs R=0.85). There is indeed a clear dependence of the ability of the model to recover tidal features with respect to the image depth: while for \mulim > 30 mag arcsec$^{-2}$ around 90\% of the tidal features are recovered, this quantity drops below 50\% for \mulim < 28 mag arcsec$^{-2}$ (see Figures \ref{fig:CM-SB}, \ref{fig:Acc-SB}). 
    \item The accuracy and purity are roughly constant at all surface brightness, hence we conclude that the model is not learning the signal-to-noise of the images, although it is evident that it impacts the classification performance, mostly in terms of completeness, as  expected.
    \item The trend with redshift is not so evident, with the larger AUC values obtained at redshift bins $z$=0.05 and $z$=0.4. The decrease of the fraction of visually identified tidal features  at higher redshifts may explain this non-intuitive result (the model is able to correctly recover the true negatives).
    \item Tidal \textit{streams} and \textit{shells} are the categories easier to identify by the model, with TPR=1 and TPR=0.87, respectively. On the other hand, \textit{mergers} and \textit{tails} reach only TPR=0.68, 0.69, respectively.
    \item When applied to real HSC images with \mulim=26 mag arcsec$^{-2}$, the performance is significantly worse than on the simulated images (AUC=0.69), even when transfer learning is applied. This is probably related to the fact that the simulated images are not extremely realistic: they have lower spatial resolution than real images and do not include background effects. Besides, they span over a wide \mulim~range, do not have a uniform redshift coverage and may not include examples of all observed morphological types and/or features.

\end{itemize}

The results presented in this work represent an important step in the development of automated tidal feature detection techniques, even if the performance  is  lower than those found in other astronomical classification tasks like separating elliptical from disk galaxies (reaching accuracy above 97\%). Tidal feature detection is a difficult task, given the  low surface brightness of the subtle structures that we wish  to detect. An additional limitation is the lack of a large, homogeneously observed training sample with certain classifications. 

One alternative  would be to use `intrinsic' labels from simulations, derived from dynamics or merger trees. The disadvantage of such labeling is that observational limitations may not always support the classification of an image in accordance with its intrinsic class. In this work, the training data is assembled from human labels. Labelling tidal features via visual inspection is not trivial and the image pre-processing has a significant impact. In addition, classifiers tend to disagree with each other quite often, as shown in \citetalias{Martin2022}. Biases in  the identification of interacting galaxies by visual classifications have also been reported in \cite{Blumenthal2020}. Therefore, it may not even be possible to achieve an accuracy similar to elliptical/spiral separation due to the much greater ambiguity of the classifications. 

We are aware of some important efforts of the scientific community towards building large and robust samples of tidal identifications, such as the detailed annotations presented in \cite{Sola2022, Bilek2020} or the on-going tidal stream survey by \cite{MartinDelgado2021}, which will certainly help to construct a robust training sample to improve the algorithms for automated detection. Approaches such as domain adaptation \citep{Ciprijanovic2022}, the use of unsupervised learning (e.g., \citealt{Sarmiento2021, Cheng2021}) or one-shot learning \citep{Chen2019} could help to overcome the lack of positive training samples currently available, but we leave these approaches for the forthcoming analysis.

The poorer performance of our tidal identification model in the real HSC-SSP images emphasizes the large dependence of deep learning algorithms on the data they are trained with. As already noted in \cite{Bottrell2019, HuertasCompany2019, Ciprijanovic2022}, the importance of using realistic simulations for training the models, including background, real noise and artifacts, is fundamental to achieving robust results with real data. This should be taken as a warning against applying deep learning models to different data domains without previously assessing their performance on the new domain.

Our results also highlight the need for deep surveys in order to construct complete samples of galaxies showing tidal interactions. Our predictions imply that we need images with \mulim > 30 mag arcsec$^{-2}$ to achieve completeness above 60$\%$, although  given the non-representative sample used for the statistical analysis presented in this work our current result is only suggestive of this requirement. For example, as already noted in \cite{Bottrell2019} and \cite{Bickley2021}, for rare objects amongst large datasets (such as the upcoming LSST) the precision, largely dependent on the assumed fraction of positive instances,  is paramount. One could increase the completeness by using larger value of P$_{\rm th}$, even if that would decrease the purity, and complement it with visual inspection of a much smaller sample of galaxies.

The challenges facing the automated detection of tidal features should not prevent us from that endeavour. The scientific return of large samples of galaxies showing tidal features is huge. The detection and characterization of these faint tidal remnants – including measurements of their
abundance, width, and shapes/morphology – probe the recent merger activity, disruption mechanisms and galaxy mass assembly. Furthermore, the characteristics of observed tidal features can constrain the global properties of the stellar and dark matter haloes \citep{Johnston1999, Sanderson2015, Bovy2016, Pearson2022} and, consequently, are a complementary way to testing cosmological and dark matter theories. These are, indeed, some of the scientific objectives of the recently approved F-ESA mission ARRAKIHS\footnote{https://www.cosmos.esa.int/documents/7423467/7423486/ESA-F2-ARRAKIHS-Phase-2-PUBLIC-v0.9.2.pdf/61b363d7-2a06-1196-5c40-c85aa90c2113?t=1667557422996} (P.I. R Guzmán), that will image 50 deg$^{2}$ of the sky per year down to an unprecedented ultra-low surface
brightness  in visible  infrared bands (31 and 30 mag arcsec$^{-2}$, respectively). The future of scientific analysis based on tidal features  is bright and promising and this work undoubtedly represents an important step forward towards understanding the requirements of an optimal automated identification of such powerful ingredient for galaxy evolution and cosmological studies.

\section*{Acknowledgements}

 We thank the anonymous referee for their suggestions, which helped to improve the clarity of the paper. HDS acknowledges support by the PID2020-115098RJ-I00 grant from MCIN/AEI/10.13039/501100011033 and from the Spanish Ministry of Science and Innovation and the European Union - NextGenerationEU through the Recovery and Resilience Facility project ICTS-MRR-2021-03-CEFCA and AdP and JAO for technical and emotional support.  I.D. acknowledges the support of the Canada Research Chair Program and the Natural Sciences and Engineering Research Council of Canada (NSERC, funding reference number RGPIN-2018-05425). FB acknowledges support from the grants PID2020-116188GA-I00 and PID2019-107427GB-C32 by the Spanish Ministry of Science and Innovation. J.H.K. acknowledges financial support from the State Research Agency (AEI-MCINN) of the Spanish Ministry of Science and Innovation under the grant "The structure and evolution of galaxies and their central regions" with reference PID2019-105602GB-I00/10.13039/501100011033, from the ACIISI, Consejer\'{i}a de Econom\'{i}a, Conocimiento y Empleo del Gobierno de Canarias and the European Regional Development Fund (ERDF) under grant with reference PROID2021010044, and from IAC project P/300724, financed by the Ministry of Science and Innovation, through the State Budget and by the Canary Islands Department of Economy, Knowledge and Employment, through the Regional Budget of the Autonomous Community. The authors gratefully acknowledge the computer resources at Artemisa, funded by the European Union ERDF and Comunitat Valenciana as well as the technical support provided by the Instituto de Física Corpuscular, IFIC (CSIC-UV).

\section*{Data Availability}

The catalogue used in this article comes from the analysis of \citetalias{Martin2022}. The code used for the deep learning algorithm will be shared upon request.





\bibliographystyle{mnras}
\bibliography{sample} 






\bsp	
\label{lastpage}
\end{document}